%% file: slac-pub-8695.tex
\documentclass[12pt]{article}
\usepackage{graphicx}

\input pubboard/babarsym
\begin{document}

\thispagestyle{empty}
\renewcommand{\thefootnote}{\fnsymbol{footnote}}

\begin{flushright}
{\small
SLAC--PUB--8695\\
\babar\--PROC--00/27\\
November 2000\\}
\end{flushright}

\vspace{.8cm}

\begin{center}
{\bf\large   
Charmless Hadronic \B\ Decays at $\babar$}

\vspace{1cm}

J. Olsen\\
Physics Department, University of Maryland\\
College Park, MD, 20742-4111\\

\medskip

(representing the \babar\ Collaboration)\\
\end{center}

\vfill

\begin{center}
{\bf\large   
Abstract }
\end{center}

\begin{quote}
We present preliminary results of several searches for rare charmless hadronic 
decays of the \B\ meson using data collected by the \babar\ detector at the 
Stanford Linear Accelerator Center's \pep2\ storage ring.  We search for
the decays $h^+h^-$, $h^+h^-h^+$, $h^+h^-\piz$, $X^0h^+$, 
and $X^0\KS$, where $h = \pi$ or $K$, and $X^0 = \eta^{\prime}$ or $\omega$.  
In a sample of $8.8$ million \BB\ decays we measure the
branching fractions:  
$\BR(\Bz\to \pip\pim) = (9.3^{+2.6}_{-2.3}$$^{+1.2}_{-1.4})\times 10^{-6}$,
$\BR(\Bz\to \Kp\pim) = (12.5^{+3.0}_{-2.6}$$^{+1.3}_{-1.7})\times 10^{-6}$,
$\BR(\Bz\to \rho^-\pip) = (49 \pm 13$$^{+6}_{-5})\times 10^{-6}$, and
$\BR(\Bu\to \eta^{\prime}K^+) = (62 \pm 18 \pm 8)\times 10^{-6}$.  We
calculate upper limits for the modes without a significant signal.
\end{quote}

\vfill

\begin{center} 
{\it Contributed to the Meeting of the Division of\\
Particles and Fields of the American Physical Society}\\
{\it Columbus, Ohio, USA}\\
{\it August 9--August 12, 2000}\\
\end{center}

\vspace{1.0cm}
\begin{center}
{\em Stanford Linear Accelerator Center, Stanford University, 
Stanford, CA 94309} \\ \vspace{0.1cm}\hrule\vspace{0.1cm}
Work supported in part by Department of Energy contract DE-AC03-76SF00515.
\end{center}

\newpage



%
\pagestyle{plain}

\section{Introduction}	        
Charmless hadronic \B\ decays will play an important role in the study of 
CP violation.  Indirect CP violation arises in $\Bz$--$\Bzb$ mixing due to 
interference between direct and mixed decays.  The CKM angle $\alpha$ 
can be measured by observing the resulting time-dependent 
asymmetry in decays to $\pi\pi$ and $\rho\pi$ final states.  Direct CP 
violation results from interference between two or more weak amplitudes and 
can arise in any decay mode where both tree and penguin contributions are 
non-negligible.  Several modes reported in this paper are ``self-tagging'',
providing efficient samples for direct CP violation searches.  Finally, 
accurate branching fraction measurements provide important tests of 
factorization models, which facilitate calculation of $\alpha$ in the presence 
of significant penguin amplitudes, and can also be used to constrain 
the CKM angle $\gamma$.~\cite{neubert00}

In this paper we summarize preliminary results of searches for the following
charmless hadronic \B\ decays:~\cite{babar1}
\begin{itemize}
 \item $\pip\pim$, $\Kp\pim$, $\Kp\Km$,
 \item $K^{*0}\pip$, $\rho^0\Kp$, $\rho^0\pip$, $\rho^-\pip$, $\Kp\pim\pip$,
 $\pip\pim\pip$,
 \item $\eta^{\prime}\Kp$, $\eta^{\prime}\KS$, $\omega h^+$, $\omega\KS$,
\end{itemize}
where charge conjugate modes are assumed throughout.  The dataset consists 
of $8.8$ million \BB\ decays collected by the \babar\ detector~\cite{babar2} 
at the \pep2\ storage ring between January and June 2000.

\section{Candidate Selection and Analysis Method}
We use only good quality tracks with a minimum transverse momentum of 
$100\mevc$ in the laboratory (LAB) frame.  Charged pions and kaons are 
identified by their energy loss ($dE/dx$) in the tracking system and the 
angle $\theta_c$ of $\check{\rm C}{\rm erenkov}$ photons produced while
traversing quartz bars~\cite{babar2}.  Neutral kaons are reconstructed in
the mode $\KS\to \pip\pim$, requiring the $\KS$ flight length to exceed 
$2\mm$ and the angle between the flight direction and momentum to be less than 
$40\mrad$.  Photon candidates are defined as calorimeter energy deposits 
unassociated with a track and having a shower shape consistent with the photon 
hypothesis.  Candidate $\piz$ and $\eta$ mesons are formed from pairs of 
photons with a minimum LAB energy of $50\mev$.  Candidate $\eta^{\prime}$ 
mesons are reconstruced in the channel $\eta\pip\pim$, where the $\eta$ mass 
is constrained to the world average value.  The $\omega$ meson is reconstructed 
in the dominant decay channel, $\omega\to \pip\pim\piz$, keeping all candidates 
within $50\mevcc$ of the known $\omega$ mass.  The $\rho$ and $K^*$ resonances 
are reconstructed in the corresponding $\pi\pi$ and $K\pi$ channels.

We select candidate \B\ mesons based on the energy-substituted mass 
$\mes$, where $\sqrt{s}/2$ is substituted for the candidate's energy, and the 
difference $\Delta E$ between the $\B$-candidate energy and $\sqrt{s}/2$.
The dominant background for all modes is continuum $q\bar{q}$ production, 
which exhibits a jet-like structure that distinguishes it from the more 
spherically symmetric \BB\ events.  To suppress this background we use the 
cosine of the angle $\theta_{\rm T}$ $(\theta_{\rm S})$ between the thrust 
(sphericity) axis of the \B\ candidate and the rest of the event, and the 
cosine of the angle $\theta_{\rm B}$ between the candidate's flight direction 
and the beam axis.  In some cases we include several event-shape variables into 
a single Fisher discriminant.

\section{Results for {\boldmath $h^+h^-$} Modes}
We select $\Bz\to h^+h^-$ candidates satisfying $5.22<\mes<5.3\gevcc$ and
$\left| \Delta E \right| < 0.420\gev$.  No explicit particle identification is 
required and the pion mass hypothesis is assumed for both tracks.  We require 
$\left|\cos{\theta_{\rm S}}\right| < 0.9$ and construct a Fisher discriminant 
${\cal F}$ from nine variables describing the momentum flow of charged and 
neutral particles around the \B\ candidate thrust axis.

\begin{figure}[tp]
\begin{center}
\includegraphics[height=6.5cm]{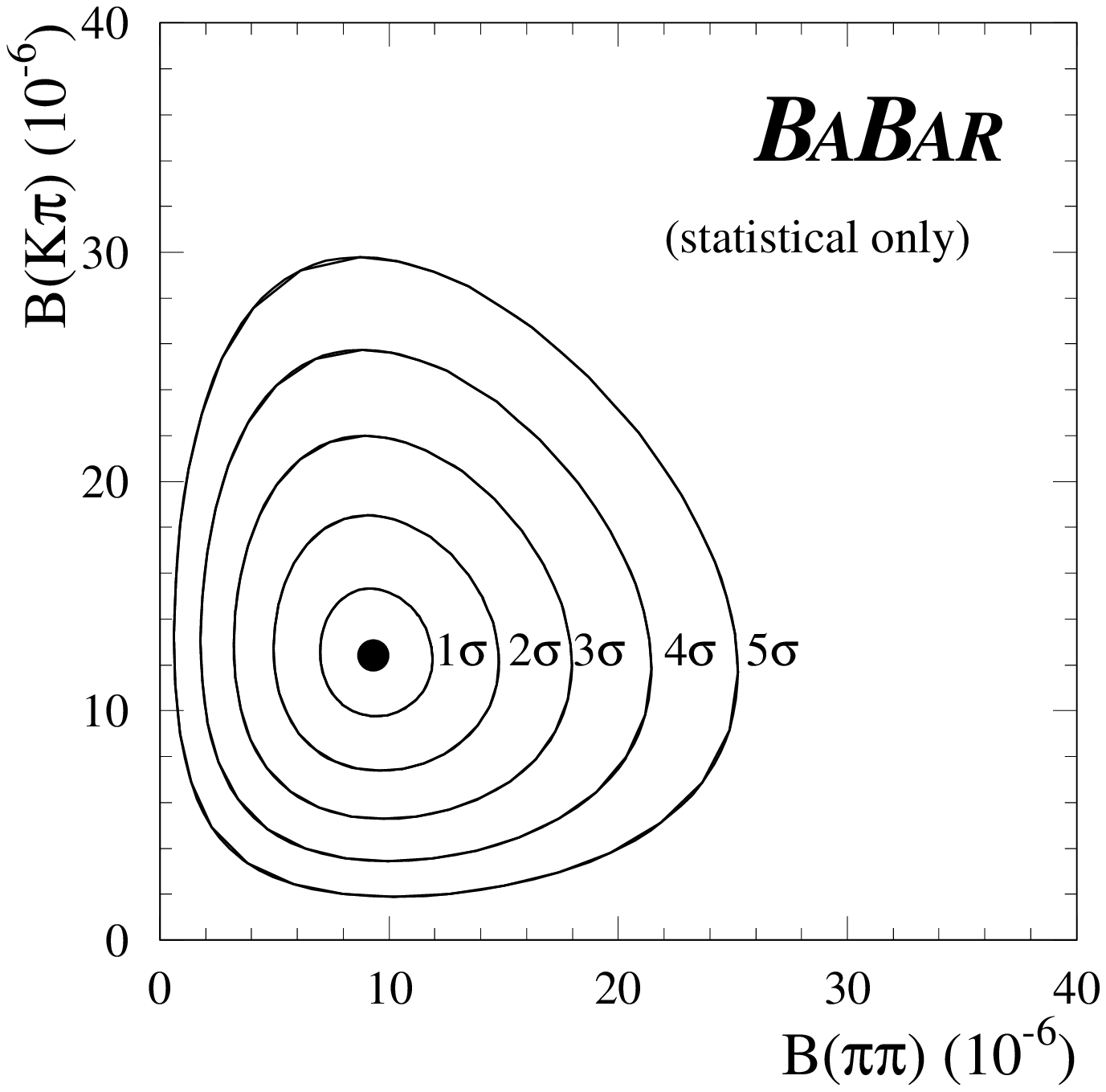}
\hspace{0.1cm}
\includegraphics[height=9.5cm,width=9.5cm]{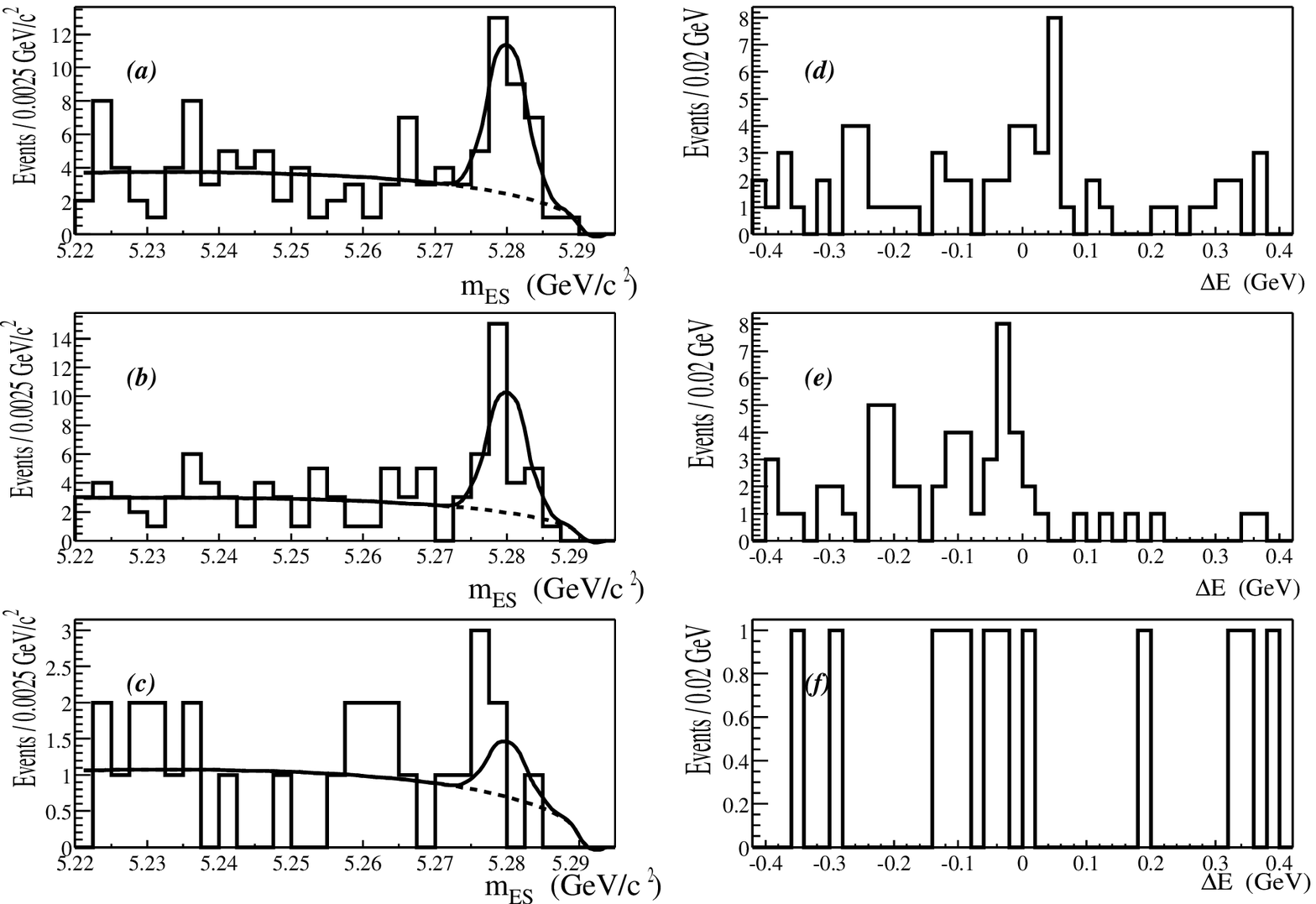}
\end{center}
\caption{Left:  The central value (filled circle) for $\BR(\Bz\to \pip\pim)$ 
and $\BR(\Bz\to \Kp\pim)$ along with the $n\sigma$ statistical contour curves 
for the global likelihood fit.  Right:  $\mes$ and $\Delta E$ for (a,d) $\pi\pi$,
(b,e) $K\pi$, and (c,f) $KK$ candidates in the cut-based analysis.}
\label{theFig}
\end{figure}

Signal yields in all three modes are determined simultaneously 
from an unbinned maximum likelihood fit incorporating $\mes$, $\Delta E$, ${\cal
F}$, and the measured $\theta_c$ for each track.  A sample of $D^*$-tagged
$D^0\to \Kp\pim$ decays is used to parameterize the $\theta_c$ distributions
for pion and kaon tracks as a function of momentum.  The $K/\pi$ separation
varies from $2$ to $8\sigma$ across the relevant momentum range.  All 
candidates in the region $-0.200 < \Delta E < 0.140\gev$ are included in the 
fit.  
We find signal yields of $N(\pi\pi) = 29^{+8}_{-7}$, $N(K\pi) = 38^{+9}_{-8}$, 
and $N(KK) = 7^{+5}_{-4}$.  As a cross-check we perform a cut-based analysis 
requiring a tighter cut on $\cos{\theta_{\rm S}}$ and additional cuts on 
$\cos{\theta_{\rm B}}$ and ${\cal F}$.  Signal yields are determined by 
applying particle identification criteria to isolate independent 
samples of candidates corresponding to each mode and then fitting the $\mes$ 
distribution in each sample.  The results are consistent with the global 
likelihood fit.  Figure~\ref{theFig} shows the global fit likelihood contour 
curves for the $\pi\pi$ and $K\pi$ modes, and the $\mes$ and $\Delta E$ 
distributions for the cut-based analysis.  The results are 
summarized in the upper section of Table~\ref{theTable}.  For the $KK$ mode we 
calculate the $90\%$ confidence level upper limit.  The dominant systematic 
errors are due to tracking efficiency and the shapes of the $\Delta E$ and 
${\cal F}$ distributions.

\section{Results for Three-body Modes}
We search for resonant three-body decays by combining a $\rho$ or $K^{*0}$ 
resonance with a charged pion or kaon.  Kaons are required to be positively
identified using $dE/dx$ and $\theta_c$ information, while tracks not
identified as kaons are assumed to be pions.  We veto any combination 
consistent with the decay $D^0\to \Km\pip$.  The selection criteria consist of 
optimized cuts on $\cos{\theta_{\rm T}}$, resonance mass, and the angle between the 
resonance daughters and the \B\ candidate momentum calculated in the rest frame 
of the vector meson.  We also explicitly search for non-resonant $\Kp\pim\pip$ 
and $\pip\pim\pip$ decays by removing all $K\pi$ and $\pi\pi$ combinations with 
invariant mass less than $2\gevcc$, and all three-body combinations consistent 
with the decay $\Bu\to \jpsi\Kp$.

We define a signal region within $6\mevcc$ of the \B\ mass in $\mes$ and
$\pm 70\mev$ in $\Delta E$.  The signal yield is determined by direct 
background subtraction, where the background in the signal region is estimated 
from the number of events in the region $5.2<\mes<5.27\gevcc$.  This method is 
cross-checked using off-resonance data.  The results are summarized in the 
middle section of Table~\ref{theTable}.  The dominant systematic errors are
due to tracking efficiency, $\piz$ efficiency, and the background 
subtraction technique.

\section{Results for Modes with {\boldmath $\eta^{\prime}$} or 
{\boldmath $\omega$}}
We search for the modes $\eta^{\prime}\Kp$, $\eta^{\prime}\KS$, $\omega h^+$,
and $\omega\KS$. For $\eta^{\prime}K$ the kaon is positively identified,
while for $\omega h^+$ the charged hadron is assumed to be a pion and the 
$\Delta E$ signal window is increased ($-0.113 < \Delta E< 0.070\gev$) to take 
into account the resulting shift in energy when the mass is mis-assigned.
The angle between the decay plane of the $\omega$ daughters and the \B\
direction in the $\omega$ rest frame is used to reduce combinatoric background.  
We require $\left|\cos{\theta_{\rm T}}\right| < 0.9$ and optimize with
respect to ${\cal F}$.  Signal yields are determined by background subtraction,
where the background is determined from off-resonance data.
The results are summarized in the lower third of Table~\ref{theTable}.  The
dominant systematic errors are the same as in the three-body analysis.

\section{Summary}
We have presented preliminary results of searches for several charmless 
hadronic \B\ decays.  Table~\ref{theTable} summarizes the results.  In all 
cases, our results are consistent with recent measurements reported by the 
CLEO~\cite{cleo} and Belle~\cite{belle} collaborations at this conference.

\begin{table}[tbp]
\caption{Branching fraction results.  Signal yields ($N_S$) for the $h^+h^-$
modes are determined from a likelihood fit, the rest are obtained by a direct
background subtraction.  Efficiencies ($\epsilon$) include intermediate branching
fractions.}
\label{theTable}
\begin{center}
\begin{tabular}{lcccc} \hline
 Mode & $N_S$ & Stat. Sig. ($\sigma$) & $\epsilon(\%)$  & $\BR\,(10^{-6})$ \\\hline
 $\Bz\to\pip\pim$         & $29 ^{+8}_{-7}$$^{+3}_{-4}$ & 5.7 & $35$  & $9.3^{+2.6}_{-2.3}$$^{+1.2}_{-1.4}$ \\
 $\Bz\to\Kp\pim$          & $38 ^{+9}_{-8}$$^{+3}_{-5}$ & 6.7 & $35$  & $12.5^{+3.0}_{-2.6}$$^{+1.3}_{-1.7}$ \\
 $\Bz\to\Kp\Km$           & $7 ^{+5}_{-4}$ ($<15$)      & 2.1 & $35$  & $<6.6$ \\
\hline
 $\Bu\to\Kstarz\pip$	  & $10.2\pm4.8$                & 2.4 & $10$  & $<28$  \\
 $\Bu\to\rho^0\Kp$	  & $10.7\pm5.1$                & 2.2 & $10$  & $<29$  \\
 $\Bu\to\Kp\pim\pip$  	  & $16.3\pm5.8$                & 3.2 & $6$   & $<54$  \\
 $\Bu\to\rho^0\pip$	  & $24.9\pm8.2$                & 3.3 & $12$  & $<39$  \\
 $\Bu\to\pip\pim\pip$ 	  & $5.4\pm5.7$                 & 0.7 & $8$   & $<22$  \\
 $\Bz\to\rho^-\pip$	  & $35.5\pm9.8$                & 4.5 & $8$   & $49\pm13^{+6}_{-5}$ \\
\hline
 $\Bu\to\eta^{\prime}\Kp$ & $12.1\pm3.7$                & 5.3 & $3$   & $62 \pm 18 \pm 8$   \\
 $\Bz\to\eta^{\prime}K^0$ & $1.4\pm1.4$                 & 1.1 & $0.6$ & $<112$ \\
 $\Bu\to\omega h^+$	  & $5.9\pm3.6$                 & 1.7 & $7.5$ & $<24$  \\
 $\Bz\to\omega K^0$	  & $-0.8\pm0.0$                & 0.0 & $2$   & $<14$  \\
 \hline
\end{tabular}
\end{center}
\end{table}

\subsection*{Acknowledgments}

\input pubboard/acknowledgements



\end{document}

%% file: pubboard/babarsym.tex
\usepackage{relsize}
\def\babar{\mbox{\slshape B\kern-0.1em{\smaller A}\kern-0.1em
    B\kern-0.1em{\smaller A\kern-0.2em R}}}

\def\piz   {\ensuremath{\pi^0}}

\def\pip   {\ensuremath{\pi^+}}
\def\pim   {\ensuremath{\pi^-}}

\def\Kbar  {\kern 0.2em\overline{\kern -0.2em K}{}}

\def\Kp    {\ensuremath{K^+}}
\def\Km    {\ensuremath{K^-}}
\def\KS    {\ensuremath{K^0_{\scriptscriptstyle S}}} 
 
\def\Kstarz  {\ensuremath{K^{*0}}}

\def\Kzb   {\ensuremath{\Kbar^0}}
\def\KzKzb {\ensuremath{K^0 \kern -0.16em \Kzb}}

\def\Dbar  {\kern 0.2em\overline{\kern -0.2em D}{}}

\def\Dzb   {\ensuremath{\Dbar^0}}
\def\DzDzb {\ensuremath{D^0 {\kern -0.16em \Dzb}}}

\def\Bz    {\ensuremath{B^0}}
\def\B     {\ensuremath{B}}
\def\Bbar  {\kern 0.18em\overline{\kern -0.18em B}{}}

\def\Bzb   {\ensuremath{\Bbar^0}}
\def\Bu    {\ensuremath{B^+}}

\def\BB    {\ensuremath{B\Bbar}} 
\def\BzBzb {\ensuremath{B^0 {\kern -0.16em \Bzb}}}

\def\jpsi  {\ensuremath{{J\mskip -3mu/\mskip -2mu\psi\mskip 2mu}}} 
\mathchardef\Upsilon="7107
\def\Y#1S{\ensuremath{\Upsilon{(#1S)}}}

\mathchardef\Deltares="7101
\mathchardef\Xi="7104
\mathchardef\Lambda="7103
\mathchardef\Sigma="7106
\mathchardef\Omega="710A
\def\Deltabar   {\kern 0.25em\overline{\kern -0.25em \Deltares}{}}
\def\Lbar {\kern 0.2em\overline{\kern -0.2em\Lambda\kern 0.05em}\kern-0.05em{}}
\def\Sigbar{\kern 0.2em\overline{\kern -0.2em \Sigma}{}}
\def\Xibar{\kern 0.2em\overline{\kern -0.2em \Xi}{}}
\def\Obar{\kern 0.2em\overline{\kern -0.2em \Omega}{}}
\def\Nbar{\kern 0.2em\overline{\kern -0.2em N}{}}
\def\Xbar{\kern 0.2em\overline{\kern -0.2em X}{}}

\def\BR{{\ensuremath{\cal B}}}

\def\mes        {\mbox{$m_{\rm ES}$}}

\def\ev   {\ensuremath{\rm \,e\kern -0.08em V}}
\def\kev  {\ensuremath{\rm \,ke\kern -0.08em V}} 
\def\mev  {\ensuremath{\rm \,Me\kern -0.08em V}} 
\def\gev  {\ensuremath{\rm \,Ge\kern -0.08em V}} 
\def\gevc {\ensuremath{{\rm \,Ge\kern -0.08em V\!/}c}} 
\def\tev  {\ensuremath{\rm \,Te\kern -0.08em V}}
\def\mevc {\ensuremath{{\rm \,Me\kern -0.08em V\!/}c}} 
\def\gevcc{\ensuremath{{\rm \,Ge\kern -0.08em V\!/}c^2}} 
\def\mevcc{\ensuremath{{\rm \,Me\kern -0.08em V\!/}c^2}}

\def\mm   {\ensuremath{\rm \,mm}}

\def\mus  {\ensuremath{\rm \,\mus}}

\def\mus        {\ensuremath{\,\mu{\rm s}}}    
\def\mrad{\ensuremath{\rm \,mr}}                

\def\gsim{{~\raise.15em\hbox{$>$}\kern-.85em
          \lower.35em\hbox{$\sim$}~}}
\def\lsim{{~\raise.15em\hbox{$<$}\kern-.85em
          \lower.35em\hbox{$\sim$}~}}

\def\to                 {\ensuremath{\rightarrow}}

\def\pep2{PEP-II}

\providecommand{\eqref}[1]{Eq.~(\ref{eq:#1})}

\def\jetset74   {\mbox{\tt Jetset \hspace{-0.5em}7.\hspace{-0.2em}4}}

%% file: pubboard/acknowledgements.tex
We are grateful for the contributions of our \pep2\ colleagues in
achieving the excellent luminosity and machine conditions
that have made this work possible.
We acknowledge support from the
Natural Sciences and Engineering Research Council (Canada),
Institute of High Energy Physics (China),
Commissariat \`a l'Energie Atomique and
Institut National de Physique Nucl\'eaire et de Physique des Particules
(France),
Bundesministerium f\"ur Bildung und Forschung
(Germany),
Istituto Nazionale di Fisica Nucleare (Italy),
The Research Council of Norway,
Ministry of Science and Technology of the Russian Federation,
Particle Physics and Astronomy Research Council (United Kingdom), the
Department of Energy (US),
and the National Science Foundation (US). In addition, individual support 
has been received from the Swiss 
National Foundation, the A. P. Sloan Foundation, the Research Corporation,
and the Alexander von Humboldt Foundation.
The visiting groups wish to thank 
SLAC for the support and kind hospitality
extended to them.